\documentclass[10pt]{article}
\usepackage{graphicx,epsf}

\newcommand{\etal}{{\sl et al.}}

\begin{document}

\title{Comment faire fondre un cristal d'\'electrons bidimensionnel sous 
champ magn\'etique}
\author{M.\ O.\ Goerbig$^{1,2}$, P.\ Lederer$^1$, and  
C.\ Morais\ Smith$^2$}

\maketitle

\noindent
{\small
\begin{center}
$^1$Laboratoire de Physique des Solides, B\^at.\,510, UPS (associ\'e au CNRS), F-91405 Orsay cedex, France.\\
$^2$D\'epartement de Physique, Universit\'e de Fribourg, P\'erolles,  CH-1700 Fribourg, Switzerland.
\end{center}}

\begin{abstract}
C'est un ph\'enom\`ene connu de la vie quotidienne : si l'on augmente la 
temp\'erature, la glace fond. Ceci est le paradigme d'une transition de phase 
solide-liquide, qu'on observe \'egalement dans d'autres mat\'eriaux : tout 
cristal fond \`a temp\'erature suffisamment \'elev\'ee. Or de telles 
transitions de phase existent aussi dans des structures moins bien connues \`a 
temp\'erature z\'ero, o\`u l'on fait varier un param\`etre physique 
diff\'erent de la temp\'erature. C'est le cas par exemple dans un syst\`eme 
d'\'electrons dont le 
mouvement est contraint dans un plan sous champ magn\'etique perpendiculaire. 
Nous avons montr\'e \cite{goerbig3} qu'une partie de ces \'electrons peut 
faire une telle transition de phase solide-liquide. Le param\`etre que l'on
varie est le champ magn\'etique m\^eme, et l'on observe un ph\'enom\`ene 
\'etrange : quand on augmente le champ magn\'etique, un cristal d'\'electrons 
fond pour former un liquide \'electronique. Contrairement \`a notre intuition 
acquise par la fusion de la glace en la chauffant, les \'electrons liquides 
forment \`a nouveau un cristal si l'on augmente davantage le champ 
magn\'etique. Une telle alternance de phases en fonction du champ a 
\'et\'e r\'ecemment observ\'ee dans des exp\'eriences par 
Eisenstein \etal~\cite{exp3} au California Institute of Technology. 
\end{abstract}

\bigskip
\noindent{\large \bf Syst\`eme d'\'electrons sans champ magn\'etique}

\medskip
\noindent
Avant de d\'ecrire le comportement des \'electrons dans un plan - on parle d'un
syst\`eme d'\'electrons bidimensionnel (2D) - sous champ magn\'etique, nous 
discutons le cas sans champ, qui est plus simple et aussi plus intuitif. Un 
tel syst\`eme se trouve par exemple dans les m\'etaux, comme le fer et le 
cuivre, o\`u les \'electrons peuvent bouger presque librement, et c'est pour
cette raison que les m\'etaux sont de bons conducteurs. Or les objet 
m\'etalliques sont normalement des objets de notre monde, qui est 
tridimensionnel (3D). O\`u trouve-t-on donc des m\'etaux 2D mis \`a part dans 
l'imagination de quelques physiciens th\'eoriciens ? L'interface entre deux 
cristaux (semi-conducteurs) de composition diff\'erente en est un exemple. 
Contrairement \`a la
partie volumique, les \'electrons sont libres \`a l'interface mais ne peuvent
pas en sortir dans la direction perpendiculaire. Ceci constitue pr\'ecis\'ement
un syst\`eme d'\'electrons 2D.

\begin{figure}
\epsfysize+6.0cm
\epsffile{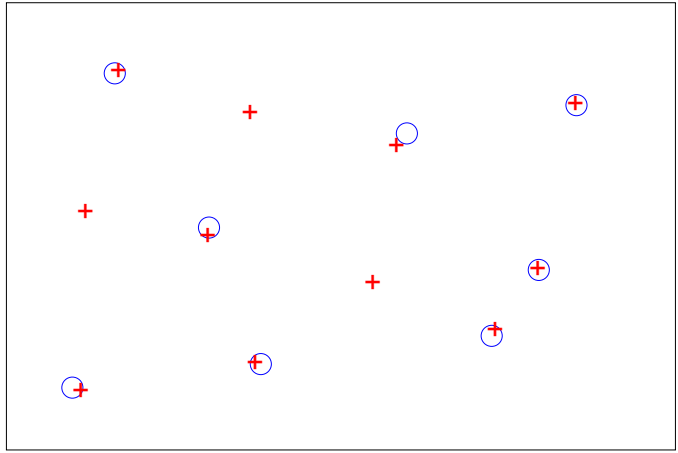}
\caption{Basse densit\'e \'electronique : les \'electrons (cercles bleus)
restent de pr\'ef\'erence sur les impuret\'es de charge positive (croix 
rouges) \`a cause de l'attraction entre particules de charge oppos\'ee. Les
\'electrons sont donc pi\'eg\'es et ne peuvent pas transporter de 
courant {\sl (isolant)}.
}
\label{fig01}
\end{figure}

Or quand on dit que ces \'electrons bougent librement dans le plan, ce n'est
qu'\`a moiti\'e vrai : \`a cause de leur charge \'electrique ils 
interagissent (a) entre eux et (b) avec les impuret\'es charg\'ees du cristal
sous-jacent, qui sont in\'evitablement pr\'esentes dans un \'echantillon 
r\'eel. Si la densit\'e d'\'electrons dans le plan est suffisamment basse, 
les \'electrons, charg\'es n\'egativement, se trouvent plut\^ot \`a la 
position des impuret\'es de charge positive, puisque des particules de charge
oppos\'ee s'attirent. Les \'electrons sont donc pi\'eg\'es par les impuret\'es.
Comme la distance moyenne entre \'electrons est grande
- ce qui signifie pr\'ecisement une basse densit\'e -, les \'electrons 
n'interagissent que peu entre eux. A cause du pi\'egeage des \'electrons,
aucun courant ne peut \^etre transport\'e, m\^eme si l'on applique une tension 
entre deux bords du plan. On a donc un {\sl isolant}; une telle
situation est esquiss\'ee dans la Fig. \ref{fig01}. 

\begin{figure}
\epsfysize+6.0cm
\epsffile{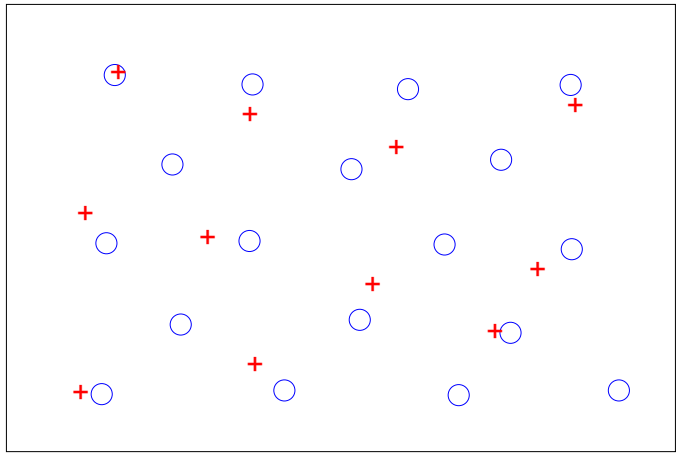}
\caption{Densit\'e \'electronique \'elev\'ee : les \'electrons se repoussent
entre eux car ils sont de m\^eme charge. Pour s'\'eviter au maximum, ils
forment une structure cristalline (cristal de Wigner). La r\'epulsion entre 
\'electrons est plus importante que leur attraction par les impuret\'es de 
charge positive. Pourtant les \'electrons se trouvent de pr\'ef\'erence au
voisinage des impuret\'es, qui d\'eforment l\'eg\`erement le cristal. Tout le 
cristal est accroch\'e aux impuret\'es, et l'on trouve \`a nouveau un
{\sl isolant}.}
\label{fig02}
\end{figure}

Si l'on augmente
la densit\'e \'electronique, on r\'eduit la distance moyenne entre \'electrons.
Il est donc impossible de n\'egliger leur r\'epulsion mutuelle, qui devient 
aussi importante, voire plus importante, que leur attraction par les 
impuret\'es. Nous n\'egligeons pour le moment l'attraction par ces impuret\'es.
Pour r\'eduire au maximum leur r\'epulsion coulombienne, c'est-\`a-dire pour
s'\'eviter au maximum, les \'electrons forment une structure cristalline. La
formation d'un tel cristal \'electronique fut propos\'ee par E.\ P.\ Wigner en 
1934 \cite{wigner}, et l'on parle d\'esormais d'un cristal de Wigner quand 
on se r\'ef\`ere \`a un cristal d'\'electrons. Quel est maintenant l'effet des 
impuret\'es, qui attirent pourtant les \'electrons d'un tel cristal de Wigner ?
Elles d\'eforment localement le cristal sans trop affecter la structure globale
si leur attraction est faible devant la r\'epulsion entre \'electrons (Fig. 
\ref{fig02}). De plus, elles accrochent le cristal et emp\^echent qu'il ne
glisse librement. On peut se repr\'esenter cette situation par une grille 
qu'on laisse glisser sur le sol. Si sa surface est bien lisse ({\sl e.g.} un 
parquet), 
la grille glisse sans probl\`eme, mais son glissement est supprim\'e si le sol 
est rugueux comme c'est le cas pour un tapis. A cause de cette suppression du 
glissement d'un cristal \'electronique par les impuret\'es, aucun transport 
\'electronique n'est possible, et le syst\`eme est \`a nouveau {\sl isolant}.

\begin{figure}
\epsfysize+6.0cm
\epsffile{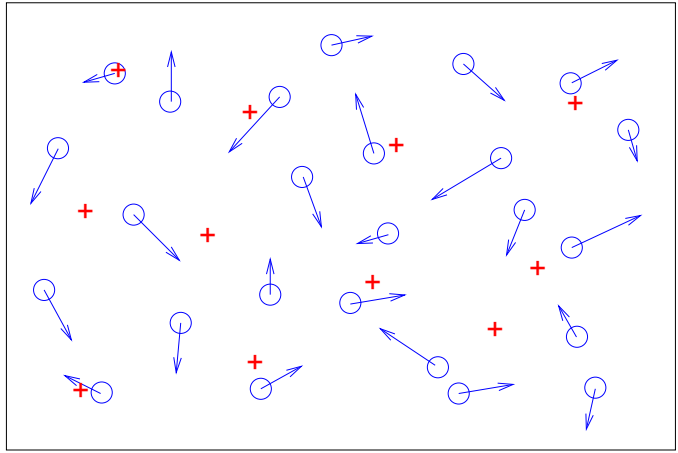}
\caption{Liquide d'\'electrons : les \'electrons bougent librement dans le 
plan (leurs vitesses sont repr\'esent\'ees par des fl\`eches bleues). A 
cause de leur r\'epulsion, la probabilit\'e de trouver deux \'electrons
\`a la m\^eme position est nulle. Pourtant il n'y a pas d'ordre cristallin :
on a un {\sl conducteur}; c'est le cas des m\'etaux.}
\label{fig03}
\end{figure}

Comment est-il donc possible de trouver des conducteurs dans la nature si les
\'electrons ne peuvent pas bouger librement dans les mat\'eriaux ? Il 
faudrait un {\sl liquide} d'\'electrons (Fig. \ref{fig03}), et non pas une 
phase cristalline ou une phase dans laquelle les \'electrons soient pi\'eg\'es 
individuellement par les impuret\'es pour r\'ealiser un conducteur. C'est 
pr\'ecisement le cas dans les m\'etaux. Un tel liquide d'\'electrons se forme
si l'on fait fondre la structure cristalline. Le m\'ecanisme de fusion le plus 
intuitif est la fusion thermique comme la fusion de la glace : 
en augmentant la temp\'erature, les \'electrons sont soumis \`a des chocs 
venant de l'ext\'erieur. Ils sont donc d\'eplac\'es de leur position initiale, 
et si l'on augmente la force des chocs (en augmentant d'avantage la 
temp\'erature), ils ne reviennent plus \`a ces positions. C'est le moment
o\`u le cristal fond. 

\begin{figure}
\epsfysize+6.0cm
\epsffile{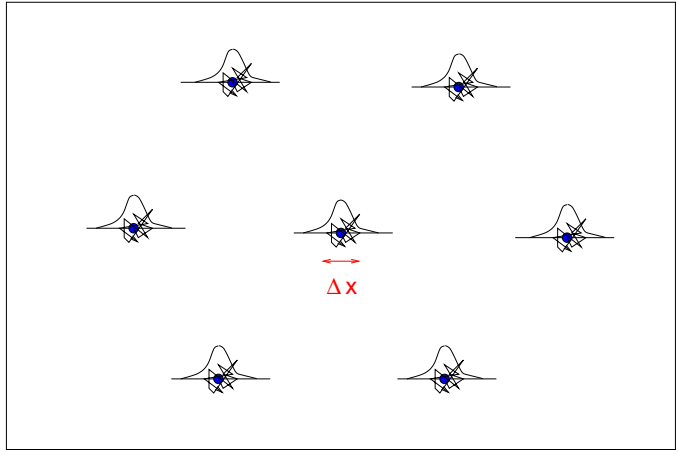}
\caption{Cristal en m\'ecanique quantique : les \'electrons peuvent bouger 
autour de leur position moyenne qui d\'efinit le cristal. Leur mouvement
est sch\'ematiquement esquiss\'e par la ligne noire. Ils sont donc 
repr\'esent\'es par une distribution de probabilit\'e appel\'ee 
{\sl fonction d'onde} (courbes noires), qui est centr\'ee autour de la 
position moyenne (points bleus) avec une largeur $\Delta x$.
En ce sens, les \'electrons ne sont plus des points mais des objets 
\'etal\'es sur une petite surface, dont l'extension dans la direction $x$ 
est pr\'ecis\'ement caract\'eris\'ee par $\Delta x$.}
\label{fig04}
\end{figure}

Mais les {\sl fluctuations thermiques} ne sont pas les seules
capables de faire fondre un cristal. Un autre mecanisme - moins connu - peut
engendrer une telle fusion, m\^eme \`a temp\'erature z\'ero, o\`u l'on 
s'attendrait \`a trouver des structures cristallines \`a cause de l'absence de 
chocs thermiques. Or la m\'ecanique quantique, qui a boulevers\'e notre 
compr\'ehension de la nature au d\'ebut du XXe si\`ecle, nous a appris
qu'on ne peut pas mesurer \`a la fois la position d'une particule et sa 
vitesse avec exactitude. Si $\Delta x$ est la barre d'erreur que l'on fait 
en mesurant sa position, on ne peut mesurer la vitesse d'une particule 
avec une pr\'ecision plus grande que $\Delta v$. Les deux barres d'erreurs 
sont reli\'ees par ce qu'on appelle la relation d'incertitude de Heisenberg
$$m\Delta x \Delta v > h,$$
o\`u $m$ est la masse de la particule et $h$ une des constantes fondamentales
de la nature {\sl (constante de Planck)}. Il est clair que le concept du 
cristal est mis en cause par 
cette incertitude : nous avons dit que dans un cristal, les particules (ici, 
les \'electrons) ont une position bien d\'etermin\'ee \`a laquelle elles 
restent fix\'ees. On devrait donc conna\^itre \`a la fois leurs positions et 
leurs vitesses, ce qui serait en contradiction avec la relation d'incertitude.
Est-il donc possible de trouver un cristal dans ces conditions ?
Il existe pourtant des cristaux dans la nature, malgr\'e la m\'ecanique 
quantique. Pour sortir de ce dilemme, il suffit de permettre aux particules
de bouger un peu autour de leurs positions dans le cristal sans trop s'en
\'eloigner. Il y a ainsi une certaine probabilit\'e de trouver
la particule \`a un endroit dans le voisinage de sa position initiale, qui joue
maintenant le r\^ole d'une position moyenne (Fig \ref{fig04}). La particule 
n'est donc plus repr\'esent\'ee par un simple point mais par une distribution 
de probabilit\'e, centr\'ee autour de la position moyenne avec une 
certaine largeur $\Delta x$, qui est exactement la barre d'erreur de la 
relation de Heisenberg. Ce mouvement quantique, qui est pr\'esent \'egalement 
\`a temp\'erature z\'ero - c'est pour cette raison qu'on parle aussi de 
{\sl fluctuations quantiques} - peut causer la formation d'un liquide 
\'electronique si la particule atteint avec une certaine probabilit\'e un 
site voisin dans le cristal. Dans ce cas, l'extension spatiale de la fonction
d'onde $\Delta x$ est de l'ordre de la distance moyenne entre \'electrons $d$,
qui est d\'etermin\'ee par la densit\'e \'electronique. On peut donc faire
fondre un cristal d'\'electrons non seulement en augmentant la temp\'erature 
{\sl (fusion thermique)} mais aussi \`a temp\'erature z\'ero si l'on 
augmente la densit\'e \'electronique {\sl (fusion quantique)}. C'est 
pr\'ecisement le cas des m\'etaux dans lesquels la densit\'e \'electronique 
est tellement \'elev\'ee qu'on ne trouve pas de phases cristallines, et c'est
pour cette raison qu'ils sont de bons conducteurs.

\bigskip
\noindent{\large \bf Syst\`eme d'\'electrons en pr\'esence d'un champ 
magn\'etique}

\begin{figure}
\epsfysize+6.0cm
\epsffile{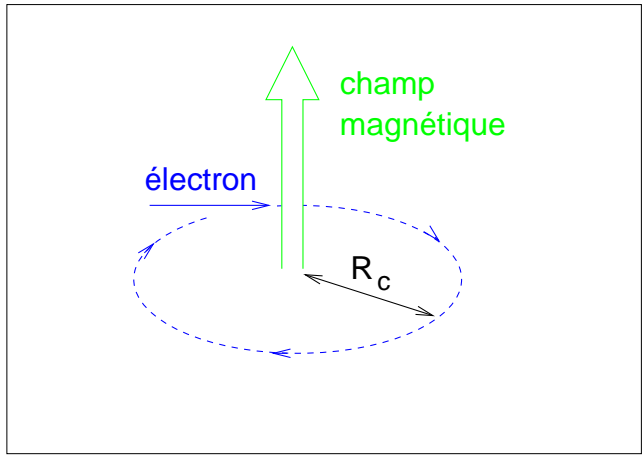}
\caption{Un \'electron, qui entre avec une vitesse finie dans un champ
magn\'etique, est d\'evi\'e sur une trajectoire circulaire. Le rayon $R_c$ 
de cette trajectoire, qui est appel\'e rayon cyclotron, est proportionnel 
\`a la vitesse de la particule. En augmentant le champ magn\'etique, on 
diminue ce rayon.}
\label{fig05}
\end{figure}

\begin{figure}
\epsfysize+6.0cm
\epsffile{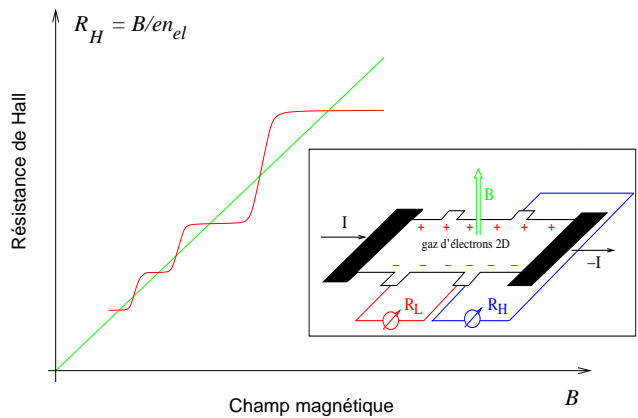}
\caption{Effet Hall : la r\'esistance tranversale (de Hall) varie 
lin\'eairement avec le champ magn\'etique. A cause de la d\'eviation
de l'\'electron dans un champ magn\'etique, il y a plus d'\'electrons au
bord inf\'erieur de l'\'echantillon qu'au bord sup\'erieur, comme esquiss\'e
dans la figure ins\'er\'ee. Cela donne lieu \`a une tension
entre les bords et par cons\'equent \`a la  r\'esistance de Hall (bleu). La
r\'esistance longitudinale (rouge) est mesur\'ee sur le m\^eme bord.
La ligne rouge montre sch\'ematiquement le comportement de la r\'esistance
de Hall \`a basse temp\'erature et \`a haut champ magn\'etique : au lieu 
d'une variation lin\'eaire avec le champ, on observe des paliers dans la 
courbe. 
Cette quantification de la r\'esistance de Hall est appel\'ee {\sl effet Hall
quantique}.}
\label{fig06}
\end{figure}

\medskip
\noindent
La situation change quand on expose le syst\`eme d'\'electrons 2D \`a un champ
magn\'etique perpendiculaire. Avant de discuter le comportement d'un ensemble
d'\'electrons avec les interactions d\'ecrites dans la partie pr\'ec\'edente,
regardons le cas d'un seul \'electron, qui entre avec une certaine 
vitesse dans le champ magn\'etique. Son mouvement, qui \'etait lin\'eaire 
en absence du champ, est maintenant soumis \`a la force de Lorentz : 
l'\'electron suit une trajectoire circulaire dans un plan perpendiculaire au
champ magn\'etique (Fig. \ref{fig05}). Le rayon, dit cyclotron, de cette
trajectoire circulaire d\'epend lin\'eairement de la vitesse de l'\'electron,
mais il est inversement proportionnel au champ magn\'etique. Cela signifie
qu'on peut diminuer ce rayon en augmentant le champ. Dans le cas du 
syst\`eme 2D dans lequel on injecte des \'electrons par le contact \`a gauche 
(courant $I$ dans la figure ins\'er\'ee dans la Fig. \ref{fig06}), cette 
d\'eviation de l'\'electron a pour effet une augmentation de la densit\'e 
\'electronique au bord inf\'erieur et une r\'eduction au bord sup\'erieur. 
Ceci cause une tension mesurable entre ces bord et par cons\'equent une 
r\'esistance $R_H$, dite de Hall. Cette r\'esistance de Hall varie 
lin\'eairement avec le champ magn\'etique (Fig. \ref{fig06}) et est 
inversement proportionnelle \`a la densit\'e \'electronique. Cet effet, 
qui fut d\'ecouvert par E.\ Hall en 1879, est encore utilis\'e aujourd'hui 
pour mesurer la densit\'e de porteurs libres dans des m\'etaux. 

En 1980, un si\`ecle apr\`es la d\'ecouverte de Hall, une exp\'erience de 
K.\ v.\ Klitzing, G.\ Dorda et M.\ Pepper a montr\'e que la r\'esistance de 
Hall \`a tr\`es basse temp\'erature ne varie pas lin\'eairement avec le 
champ magn\'etique mais que quelques valeurs de la r\'esistance sont 
sp\'eciales \cite{klitzing} : \`a certains champs, la r\'esistance reste 
constante quand on varie l\'eg\`erement le champ autour de cette valeur, 
ce qui donne lieu \`a des paliers dans la r\'esistance de Hall autour de la 
courbe classique (courbe rouge dans la Fig. \ref{fig06}). La r\'esistance est
donc {\sl quantifi\'ee}, et c'est effectivement une cons\'equence de 
cette \'etrange m\'ecanique quantique dont nous avons parl\'e dans la section
pr\'ec\'edente. La d\'ecouverte de cet effet Hall quantique fut recompens\'ee 
par le Prix Nobel, attribu\'e \`a K.\ v.\ Klitzing en 1985.

\begin{figure}
\epsfysize+5.0cm
\epsffile{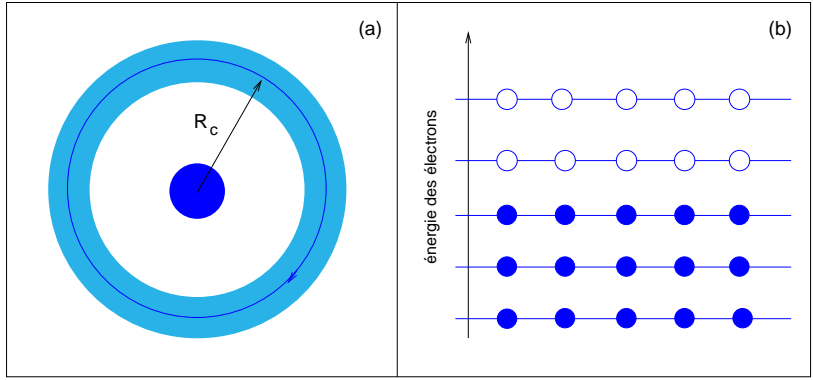}
\caption{{\sl (a)} fonction d'onde d'un \'electron dans un champ 
magn\'etique. On trouve l'\'electron avec la plus grande probabilit\'e (bleu 
clair) dans le voisinage de sa trajectoire classique (bleu fonc\'e). Aussi 
le centre de la trajectoire est-il \'etal\'e sur une surface (bleu fonc\'e).
{\sl (b)} \'energie des \'electrons sous champ magn\'etique. On 
trouve des niveaux d'\'energie. Les \'energies entre ces niveaux sont 
interdites aux \'electrons mais chaque niveau contient $N_B$ places. Chacune 
de ces places - les {\sl \'etats} - peut \^etre occup\'ee par un
\'electron (cercles bleus) ou rester vide (cercles blancs). Ici, nous avons
montr\'e le cas o\`u un nombre entier de niveaux sont compl\`etement remplis
tandis que les autres sont vides.}
\label{fig07}
\end{figure}

Comment peut-on comprendre cet \'etrange comportement d'\'electrons dans le
monde quantique - sans trop entrer dans la th\'eorie compliqu\'ee de la
m\'ecanique quantique ? Nous avons d\'ej\`a mentionn\'e que l'\'electron est 
d\'ecrit par une fonction d'onde, qui donne la probabilit\'e de le trouver 
\`a une certaine position. En pr\'esence d'un champ magn\'etique, cette 
fonction d'onde a la forme d'un anneau (Fig. \ref{fig07}) de rayon $R_c$. 
C'est une particularit\'e de la m\'ecanique quantique que ce rayon ne peut
plus prendre n'importe quelle valeur mais seulement des valeurs 
$R_c=l_B\sqrt{2n+1}$, o\`u $n$ est un entier. La longueur minimale $l_B$ 
joue le r\^ole de la constante fondamentale $h$ dans la relation 
d'incertitude, introduite en haut. Cela ressemble au cas d'un \'electron
qui tourne dans un atome autour du noyau sur des orbites dont le rayon ne 
peut prendre que des valeurs pr\'ecises. Ce cas - le mod\`ele de Bohr - est
peut-\^etre plus familier pour le lecteur. Cette quantification du rayon 
d'orbite de l'\'electron a aussi pour cons\'equence que son \'energie 
peut prendre uniquement certaines valeurs pr\'ecises. Ceci est repr\'esent\'e 
par des niveaux d'\'energie (Fig. \ref{fig07}, droite); toute valeur entre ces
\'energies est interdite aux \'electrons. Or chaque niveau contient un certain
nombre ($N_B$) de places, qu'on appelle les {\sl \'etats}, dont chacune peut 
\^etre occup\'ee par un \'electron. Il est \'evident que le remplissage de 
ces niveaux est d\'etermin\'e par le rapport $N_{el}/N_B$ o\`u $N_{el}$ est 
le nombre d'\'electrons dans le syst\`eme. 

Regardons d'abord le cas o\`u $n$ niveaux sont compl\`etement remplis, et les
niveaux sup\'erieurs restent vides. Cela ressemble \`a une situation chimique :
dans les gaz nobles, comme l'h\'elium ($He$), le n\'eon ($Ne$) ou l'argone 
($Ar$), les plus basses couches - elles ne sont pas autre chose que nos 
niveaux d'\'energie - sont compl\`etement remplies. Dans cette situation,
les \'electrons peuvent \^etre trait\'es comme inertes, {\sl i.e.} m\^eme 
s'ils sont pr\'esents, ils ne r\'epondent pas au monde ext\'erieur. C'est pour 
cette raison que les gaz nobles r\'eagissent si peu avec d'autres 
\'el\'ements. La situation est exactement la m\^eme dans notre cas 
d'\'electrons soumis \`a un champ magn\'etique : les \'electrons dans un 
niveau compl\`etement rempli n'influencent pas le comportement des \'electrons
dans un autre niveau. Commen\c cons maintenant \`a remplir partiellement 
un niveau sup\'erieur (ce qui peut \^etre effectu\'e en abaissant le champ 
magn\'etique car celui-ci d\'efinit le nombre de places par niveau). M\^eme 
si ces \'electrons n'interagissent pas avec les \'electrons dans les niveaux
inf\'erieurs, ils interagissent \`a nouveau entre eux avec les impuret\'es 
de l'\'echantillon. Le cas d'un syst\`eme d'\'electrons sans champ se reproduit
seulement pour les \'electrons du dernier niveau : \`a basse densit\'e ces
\'electrons sont pi\'eg\'es par les impuret\'es, et si l'on augmente leur
nombre, ils forment un cristal \`a cause de leur r\'epulsion (Figs. \ref{fig01}
et \ref{fig02}). Dans les deux cas, nous retrouvons un comportement isolant 
des \'electrons dans le niveau partiellement rempli, comme nous l'avons 
discut\'e
dans la section pr\'ec\'edente. Quand on varie le champ magn\'etique 
autour d'une valeur qui correspond \`a un nombre entier de niveaux 
compl\`etement remplis, les propri\'et\'es de transport ne changent donc pas 
parce que les \'electrons qui peuplent un niveau sup\'erieur sont isolants et
ne contribuent pas au transport. Or les propri\'etes de transport sont 
pr\'ecis\'ement mesur\'ees par la r\'esistance, et c'est pour cette raison
que la r\'esistance de Hall ne change pas si l'on varie le champ magn\'etique
autour de cette valeur. {\sl Ceci donne lieu \`a la formation d'un palier 
dans la r\'esistance de Hall, qui est donc li\'ee au comportement isolant 
d'\'electrons dans un niveau partiellement rempli - l'effet Hall quantique} 
(Fig. \ref{fig06}).

\begin{figure}
\epsfysize+6.0cm
\epsffile{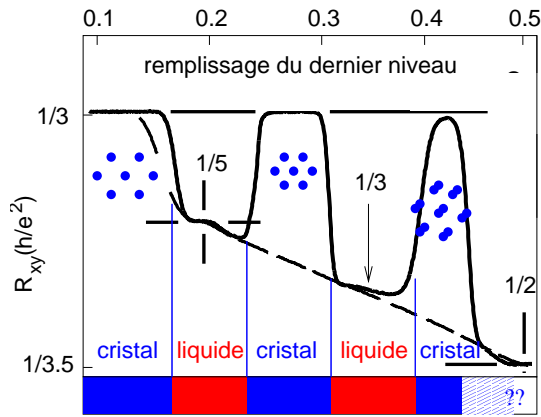}
\caption{Alternance de phases \'electroniques cristallines (bleues) et liquides
(rouges). La courbe montre la r\'esistance de Hall en fonction du remplissage 
du dernier niveau mesur\'ee par Eisenstein {\sl et al.} (Ref. \cite{exp3}). Le
palier indique que les \'electrons sont isolants. Or ce palier est interrompu
l\`a o\`u la r\'esistance est abaiss\'ee. Ceci correspond \`a la formation 
d'un liquide d'\'electrons en accord avec nos r\'esultats th\'eoriques 
\cite{goerbig3}. A un remplissage $\sim 0.42$, le cristal \'electronique
contient deux \'electrons par site.}
\label{fig08}
\end{figure}

Or nous avons vu qu'il y a aussi des phases liquides d'\'electrons, et l'on 
peut s'attendre \'egalement \`a les trouver dans un niveau partiellement 
rempli. On trouve effectivement de telles phases liquides dans le 
premier niveau excit\'e, {\sl i.e.} quand le plus bas niveau d'\'energie 
est compl\`etement rempli. Pour d\'eterminer th\'eoriquement quelle phase est 
r\'ealis\'ee \`a quel remplissage, nous avons compar\'e l'\'energie des 
phases cristallines \`a celle des liquides; c'est la phase avec la plus basse 
\'energie qui gagne \cite{goerbig3}. Nos r\'esultats sont sch\'ematiquement 
montr\'es dans la figure \ref{fig08} : on trouve des phases liquides dans le
voisinage d'un remplissage $1/5$ et $1/3$ du dernier niveau (partie rouge). 
Ces phases sont entour\'ees par des phases cristallines (partie bleue) dont
celle autour d'un remplissage $\sim 0.42$ est particuli\`ere : il s'agit d'un
cristal avec {\sl deux} \'electrons par site. Nous soulignons un effet 
inattendu : en augmentant le remplissage et donc la densit\'e \'electronique 
du dernier niveau, le cristal \'electronique fond d'abord, mais il est 
reconstitu\'e
si l'on augmente davantage la densit\'e. Ceci se rep\`ete une fois. Pour
se rendre compte \`a quel point cet effet est \'etrange, il faut s'imaginer
la glace qui fond \`a temp\'erature $T=0^0$C et qui reg\`elerait si l'on 
augmentait encore la temp\'erature, disons \`a $10^0$C ! Or une
telle situation arrive dans le cas d'un cristal \'electronique \`a cause 
des \'etranget\'es de la m\'ecanique quantique. Cette alternance de phases
liquides et cristallines nous permet de comprendre une exp\'erience r\'ecente
d'Eisenstein {\sl et al.} \cite{exp3}, qui ont mesur\'e la r\'esistance de 
Hall dans ce r\'egime de remplissage (courbe dans la Fig. \ref{fig08}). 
Ils ont montr\'e qu'\`a faible 
remplissage, on trouve le palier dans cette r\'esistance qui indique le 
comportement isolant des \'electrons dans le dernier niveau, comme nous l'avons
discut\'e plus haut dans cette section. Autour des remplissages, pour lesquels
nos calculs indiquent l'existence d'une phase liquide, cette r\'esistance est 
abaiss\'ee : les \'electrons liquides sont donc conducteurs, conform\'ement
\`a nos attentes. Autour de demi-remplissage, les \'electrons sont \`a nouveau
liquides mais l'origine de cette phase n'est pas encore compl\`etement 
expliqu\'ee.

\bigskip
En conclusion, nous avons montr\'e qu'il y a une 
alternance de phases cristallines et liquides en fonction du remplissage du 
dernier niveau 
d'\'energie si le plus bas niveau est compl\`etement rempli : en augmentant
la densit\'e d'\'electrons dans ce dernier niveau, on trouve d'abord une 
fusion du cristal \'electronique. Le liquide est form\'e autour d'un 
remplissage $1/5$ avant qu'on ne retrouve le cristal \`a plus haute densit\'e.
Ce serait aussi \'etrange que si l'eau, quand on augmente la temp\'erature,
se mettait \`a geler au lieu de bouillir. Ce ph\'enom\`ene est 
caract\'eristique de la m\'ecanique quantique.
Autour d'un remplissage de $1/3$, on retrouve une phase liquide
qui se cristallise \`a nouveau autour de $0.41$, mais ce cristal 
contient deux
\'electrons par site. Dans ce sc\'enario, nous avons expliqu\'e des
exp\'eriences r\'ecentes. 

\bigskip
Nous remercions Mathilde L\'ev\^eque pour sa contribution \`a l'essai de 
rendre compr\'ehensible notre travail pour d'\'eventuels lecteurs qui n'ont 
pas de formation scientifique sup\'erieure.

\newpage

\centerline{\large \bf Dictionnaire}

\bigskip
\noindent
{\sl Propri\'et\'es de conduction :}

\medskip
\noindent
(a) r\'esistance : grandeur physique par laquelle on caract\'erise les 
propri\'et\'es de conduction \'electrique des mat\'eriaux. Elle est le rapport
entre la tension $U$ et le courant $I$, $R=U/I$. Pour un {\sl conducteur}, 
elle 
prend une valeur finie, tandis qu'elle devient infinie pour un {\sl isolant}.\\
(b) r\'esistance de Hall (ou : transversale) : r\'esistance qu'on mesure \`a
travers la tension entre deux bords oppos\'es d'un \'echantillon rectangulaire
quand on laisse passer un courant
par les deux autres bords (figure ins\'er\'ee dans Fig. \ref{fig06}).\\
(c) effet Hall classique : la r\'esistance de Hall varie lin\'eairement avec 
le champ magn\'etique (ligne verte dans Fig. \ref{fig06}) et elle est 
inversement proportionnelle \`a la densit\'e \'electronique.\\
(d) effet Hall quantique : apparition de paliers dans la r\'esistance de Hall 
\`a basse temp\'erature (ligne rouge dans Fig. \ref{fig06}); il est d\^u au 
comportement isolant des \'electrons dans le dernier niveau partiellement 
rempli.

\bigskip
\noindent
{\sl Phases \'electroniques et transitions de phase: }

\medskip
\noindent
Comme pour d'autres particules, on trouve les \'electrons dans de diff\'erents
\'etats qu'on appelle {\sl phases}. \\
(a) phase d\'esordonn\'ee : les \'electrons se trouvent dans un \'etat o\`u il
n'y a pas d'ordre. Si l'on conna\^it la position d'un \'electron, on ne peut 
rien dire sur la position des autres. En pr\'esence
d'impuret\'es de charge positive, qui attirent les \'electrons, ils sont 
pi\'eg\'es et ne peuvent pas transporter de courant {\sl (isolant)}.\\
(b) phase cristalline (ou solide) : les \'electrons forment un cristal \`a 
cause de leur r\'epulsion mutuelle. Si l'on conna\^it la position d'un 
\'electron, on peut d\'eduire la position des autres dans le cas id\'eal. 
En pr\'esence
d'impuret\'es charg\'ees, ce cristal est accroch\'e et donc {\sl isolant}.\\
(c) phase liquide : les \'electrons bougent librement sans qu'il y ait un 
ordre qui permette de d\'eduire la position de toutes les particules \`a partir
de la position d'une seule. A cause du mouvement libre des \'electrons, cette 
phase est un {\sl conducteur}.\\
(d) fusion thermique : transition de phase qui a lieu quand un cristal fond
et forme un liquide quand on augmente la temp\'erature.\\
(e) fusion quantique : m\^eme transition \`a temp\'erature z\'ero quand on fait
varier un param\`etre physique diff\'erent de la temp\'erature.

\bigskip
\noindent
{\sl Rayon cyclotron $R_c$: }

\medskip
\noindent
Rayon de la trajectoire circulaire d'une particule charg\'ee dans un champ 
magn\'etique; il est proportionnel \`a la vitesse de la particule et 
inversement
proportionnel au champ magn\'etique. (a) {\sl classique :} $R_c$ peut prendre
n'importe quelle valeur; (b) {\sl quantique :} $R_c=l_B\sqrt{2n+1}$ ne peut 
prendre que des valeurs avec $n$ entier (en termes d'une longueur minimale 
$l_B=\sqrt{h/2\pi e B}$).

\bigskip
\noindent
{\sl Relation d'incertitude de Heisenberg :}

\medskip
\noindent
Relation fondamentale de la m\'ecanique quantique qui constate qu'on ne peut 
pas mesurer avec exactitude \`a la fois la position et la vitesse d'une 
particule. (Il y a d'autres grandeurs physiques dont la mesure est contrainte 
par une telle relation.)

\bigskip
\noindent
{\sl Niveaux d'\'energie :}

\medskip
\noindent
En m\'ecanique quantique, l'\'energie d'un syst\`eme physique 
(ici : syst\`eme d'\'electrons) ne
peut souvent prendre que des valeurs discr\`etes {\sl (quantification de 
l'\'energie)}. Ces valeurs d\'efinissent les 
{\sl niveaux d'\'energie}, qui peuvent contenir plusieurs places 
{\sl (\'etats)}. Ces places sont soit vides, soit remplies chacune exactement 
par un \'electron {\sl (principe de Pauli)}. C'est pr\'ecis\'ement le 
cas des \'electrons dans un champ magn\'etique $B$, dont l'\'energie est 
quantifi\'ee en niveaux \'equidistants qui contiennent chacun $N_B\propto B$ 
places. Le {\sl remplissage} de ces niveaux est d\'etermin\'e par le rapport
$N_{el}/N_B$, o\`u $N_{el}$ est le nombre total d'\'electrons dans le 
syst\`eme.

\newpage

\end{document}